\documentstyle[aps,prb,multicol,epsf]{revtex}

\begin{document}

\title{
Effects of disorder on 
two coupled Hubbard chains at half-filling
}

\author{Satoshi Fujimoto}
\address{Department of Physics, Kyoto University, Kyoto 606, Japan}

\author{Norio Kawakami}
\address{Department of Applied Physics,
Osaka University, Suita, Osaka 565, Japan}

\date{\today}

\maketitle
     
\begin{abstract}
We investigate the effects of quenched disorder on two chain 
Hubbard models at half-filling
by using bosonization and renormalization group methods.
It is found that the sufficiently strong 
forward scattering due to impurities 
and the random gauge field, 
which is generated by impurity backward scattering, 
destroy the charge gaps as well as
the spin gaps. Random backward scattering due to impurities
then drives the resulting massless phase to 
the Anderson localization phase.
For intermediate strength of random forward scattering, however,
the spin gaps still survive, and only one of the charge gaps 
is collapsed. In this parameter region, 
one of the charge degrees of freedom 
is in the Anderson localized state, while the other one is still 
in the massive state.
\end{abstract}

\begin{multicols}{2}

\section{Introduction}

Coupled chain systems with ladder structure 
have attracted much current interest in the study of 
strongly correlated electron systems.
It has been proposed 
that these systems could show a superconducting state
in the metallic phase.\cite{drs,grs,nws}
Such ladder systems have been realized experimentally in 
${\rm Sr_{n-1}Cu_{n+1}O_{2n}}$ and ${\rm VO_2P_2O_7}$.\cite{sr,vo}
More recently it has been reported that 
hole-doped ${\rm Sr_{n-1}Cu_{n+1}O_{2n}}$ systems 
actually undergo the transition
to the superconductivity under pressure.\cite{aki}
As microscopic models for such systems, two coupled Hubbard chains 
and two coupled $t$-$J$ chains have been extensively 
studied.\cite{drs,grs,nws,fin,fab,ima,ttr,kv,fuji1,nao,shu,baf}
The ground state of these models at half-filling is the Mott
insulator with spin gap.
When holes are doped into these systems, a metallic state with spin 
gap is realized. The resulting massless charge mode belongs to a 
class of the Tomonaga-Luttinger (TL) liquid
(sometimes referred to as the Luther-Emery class), 
and has the enhanced fluctuation toward superconductivity
This superconducting state may be related with
that found in ${\rm Sr_{n-1}Cu_{n+1}O_{2n}}$.

Such ladder systems are quasi-one-dimensional, so that they
may be quite sensitive to random potentials.
It is thus important to investigate the 
effects of quenched disorder on
two coupled Hubbard chains.
The case away from half-filling was studied 
by Orignac and Giamarchi before.\cite{og}
However the case at half-filling has not been considered so far.
In this case, we can systematically study 
how the Mott transition competes with the Anderson localization
when the disorder is introduced to the system.
This issue was previously addressed 
by the present authors, but for one-dimensional (1D)
interacting electron systems.\cite{fuji}(see also refs.18 and 19)
In contrast to such single chain models, the Mott 
insulating state for
two coupled Hubbard chains at half-filling has the 
spin gap as well as the
charge gap.  Since the randomness affects not only the charge gap 
but also  the spin gap, it is expected that the presence 
of the spin gap may bring about novel properties, 
which have not been 
observed in the single chain model,
for the competition between 
the Mott transition and the Anderson localization 
 

In this paper,
we systematically study the effects of disorder on the 
two coupled Hubbard chains at half-filling.
By exploiting bosonization and renormalization group methods,
we discuss how the introduction of disorder drives the system to     
the Anderson localized state.
In particular, we focus on the role played by the 
spin gap for the competition between the 
Mott insulator and the Anderson localization.

We deal with two types of quenched disorder:
random impurity potentials within each band, and 
random hopping between two bands.
We first discuss the effects of the former type of disorder.
As was pointed out in ref. 17,
the backward scattering due to impurities generates random gauge 
field coupled with electron currents.  We find that
the resulting random gauge fields play an essential role 
for destroying a charge gap in the coupled-chain systems.
At half-filling, sufficiently strong random forward scattering 
due to impurities and random gauge fields 
destroy charge gaps as well as spin gaps, and the 
resulting massless charge modes are localized 
by impurity  backward scattering.
However, for intermediate strength of random forward scattering,
only one of the charge gaps is collapsed, and 
other modes are still gapful.
In this case, we find a parameter region where
one charge degrees of freedom is in the Anderson localized state,
while the other one is still in the massive state.

In the case of random hopping between two bands,
the analysis is more difficult, because two charge modes are
mixed by random hopping.
However introducing a unitary transformation,
we can avoid this difficulty and find the same low-energy 
fixed point as the case without random interband hopping. 

The organization of the paper is as follows:
In Sec. II, we introduce the model for the coupled Hubbard
chains, and apply bosonization methods
to obtain low-energy effective theory. For later convenience,
we also summarize the results known for a clean 
system without randomness. 
In Sec. III, we derive the scaling equations for 
a disordered system at half-filling. 
Sec. IV and V  are devoted to the discussions 
about the fixed point properties in the half-filling case.
In particular, we discuss in detail how the competition
between the Mott transition  and the Anderson transition
occurs when the random potentials are introduced.
 In Sec. VI, we further take into account 
the effects of random hopping  between two bands.
Conclusion is given in Sec. VII.

\section{Model and bosonization method: summary of the results 
for clean systems}

We first summarize the results for a clean system
without random potentials, and then investigate the effects of 
disorder on the resulting fixed points. 
Our model Hamiltonian in the clean case is given by,
\begin{eqnarray}
H&=&-t\sum_{i,s,\alpha}
c^{\dagger}_{i s\alpha}c_{i+1 s\alpha}+h.c.
+U\sum_{i, \alpha}n_{i\uparrow\alpha}n_{i\downarrow\alpha}      
\nonumber \\
&&-t_{\perp}\sum_{i, s}[c^{\dagger}_{i s 1}c_{i s 2}+h.c.],
\label{model}
\end{eqnarray} 
where $c_{i s\alpha}$($c^{\dagger}_{i s\alpha}$) is 
an annihilation (creation) operator for electrons with spin 
$s=\uparrow \downarrow$ and chain-index $\alpha=1,2$ at a site $i$, and 
$n_{i s\alpha}=c^{\dagger}_{i s\alpha}c_{i s\alpha}$.
The last term is the hopping term between two chains.
The kinetic energy part is diagonalized in terms of a new basis,
\begin{eqnarray}
\psi_{s 1}(k)&=&\frac{c_{s 1}(k)-c_{s 2}(k)}{\sqrt{2}}, \\
\psi_{s 2}(k)&=&\frac{c_{s 1}(k)+c_{s 2}(k)}{\sqrt{2}}.
\end{eqnarray}
The dispersion relations in  this basis are given by 
$\varepsilon_{1 k}=-2t\cos k + t_{\perp}$ and
$\varepsilon_{2 k}=-2t\cos k - t_{\perp}$, resulting in
the two decoupled bands. For $t_{\perp}>2t$, there exists a band 
gap between these two bands.
In the following we restrict our 
arguments to the interesting case of $t_{\perp}<2t$,
where the chemical potential may cross the two bands 
both at and near half-filling. 
We apply abelian bosonization methods to this model
to obtain the scaling equations.\cite{eme,hal}
In the presence of spin degrees of freedom, non-abelian 
bosonization is generally more convenient for preserving 
SU(2) symmetry.\cite{witten,aff1}
However, in our model the backward scattering due to impurities
is expressed in terms  of the operators which are not included in
the operator content of SU(2) Wess-Zumino-Witten model. 
Therefore we exploit conventional abelian bosonization methods
by carefully dealing with SU(2) symmetry 
of spin degrees of freedom.
Passing to the continuum limit, 
we linearize the dispersions around the Fermi momenta
$k_{\rm F1,2}=\vert \cos^{-1}((\pm t_{\perp}-\mu)/2t)\vert$ 
($\mu$: chemical potential), and
divide the electron operators into left- and right-going parts,
$\psi_{s 1}\sim \exp(i\sqrt{4\pi}\phi_{s 1 L}(x)+ik_{\rm F1}x)
+\exp(-i\sqrt{4\pi}\phi_{s 1 R}(x)-ik_{\rm F1}x)$, etc.
Defining the new boson phase fields,
\begin{eqnarray}
\phi_{+\rho}(x)&=&\frac{\phi_{\uparrow 1}(x)+\phi_{\downarrow 1}(x)
+\phi_{\uparrow 2}(x)+\phi_{\downarrow 2}(x)}{2}, \\
\phi_{-\rho}(x)&=&\frac{\phi_{\uparrow 1}(x)+\phi_{\downarrow 1}(x)
-\phi_{\uparrow 2}(x)-\phi_{\downarrow 2}(x)}{2}, \\
\phi_{+\sigma}(x)&=&\frac{\phi_{\uparrow 1}(x)-\phi_{\downarrow 1}(x)
+\phi_{\uparrow 2}(x)-\phi_{\downarrow 2}(x)}{2}, \\
\phi_{-\sigma}(x)&=&\frac{\phi_{\uparrow 1}(x)-\phi_{\downarrow 1}(x)
-\phi_{\uparrow 2}(x)+\phi_{\downarrow 2}(x)}{2}, 
\end{eqnarray}
with $\phi_{s 1(2)}(x)\equiv\phi_{s 1(2) L}(x)+\phi_{s 1(2) R}(x)$ 
($s=\uparrow,\downarrow$),
and their canonical conjugate momentum fields, 
$\Pi_{+\rho}(x)=\partial_x\theta_{+\rho}(x)
\equiv\partial_x(\phi_{+\rho L}-\phi_{+\rho R})$, etc.,
we write down the low-energy effective Hamiltonian,
\begin{eqnarray}
H&=&H_0+H_1+H_2, \\
H_0&=&\sum_{a=\pm,\nu=\rho,\sigma}
\int dx\bigl[\frac{v_{a\nu}}{2K_{a\nu}}
(\partial_x\phi_{a\nu}(x))^2 \nonumber \\
&& \qquad\qquad +\frac{v_{a\nu} K_{a\nu}}{2}
\Pi_{a\nu}^2(x)\bigr], \\
H_1&=&U\int\frac{dx}{\alpha}\cos(\sqrt{8\pi}\phi_{1\rho}+\delta_1 x) 
\nonumber \\
&&+U\int\frac{dx}{\alpha}\cos(\sqrt{8\pi}\phi_{2\rho}+\delta_2 x), \\
H_2&=&U_1\int \frac{dx}{\alpha}\cos\sqrt{4\pi}\theta_{-\rho}
\cos\sqrt{4\pi}\phi_{-\sigma} \nonumber \\
&&+U_2\int \frac{dx}{\alpha}
\cos\sqrt{4\pi}\theta_{-\rho}\cos\sqrt{4\pi}\phi_{+\sigma}  
\nonumber \\
&&+U_3\int \frac{dx}{\alpha}\cos\sqrt{4\pi}\theta_{-\rho}
\cos(\sqrt{4\pi}\phi_{+\rho}+\delta x)  \nonumber \\
&&+U_4\int \frac{dx}{\alpha}
\cos\sqrt{4\pi}\theta_{-\sigma}\cos\sqrt{4\pi}\phi_{+\sigma} 
\nonumber \\
&&+U_5\int \frac{dx}{\alpha}\cos\sqrt{4\pi}\theta_{-\sigma}
\cos(\sqrt{4\pi}\phi_{+\rho}+\delta x) \nonumber \\
&&+U_6\int \frac{dx}{\alpha}
\cos(\sqrt{4\pi}\phi_{+\rho}+\delta x)
\cos\sqrt{4\pi}\phi_{-\sigma} \nonumber \\
&&+U_7\int \frac{dx}{\alpha}\cos\sqrt{4\pi}\phi_{-\sigma}
\cos\sqrt{4\pi}\phi_{+\sigma} \nonumber \\
&&+U_8\int \frac{dx}{\alpha}
\cos(\sqrt{4\pi}\phi_{+\rho}+\delta x)
\cos\sqrt{4\pi}\phi_{+\sigma} \nonumber \\
&&+U_9\int \frac{dx}{\alpha}\cos\sqrt{4\pi}\theta_{-\sigma}
\cos\sqrt{4\pi}\theta_{-\rho}, \label{int1}
\end{eqnarray}
where $\delta_{1,2}=4k_{{\rm F} 1,2}-2\pi$, and 
$\delta=2(k_{{\rm F}1}+k_{{\rm F}2})-2\pi$.
$H_1$ and $H_2$ are, respectively, the umklapp scattering terms 
within each band and the other scattering terms of 
electron-electron interaction.
For on-site Coulomb interaction, initially $U_i=U$ for $i=1\sim7$
and $U_8=U_9=0$.
We have dropped the terms
which include an oscillating factor 
$\exp(\pm 2i(k_{\rm F1}-k_{\rm F2})x)$,
because these terms are irrelevant in the long wave-length limit.
We have also omitted the forward scattering terms of              
electron-electron
interaction, which just renormalize the TL parameters.

We first consider the case of half-filling
$\delta=0$. Then $H_1$-term is irrelevant and can be discarded
after the renormalization procedure 
because of the oscillating factors.
In general, the above Hamiltonian produces 
the mass gaps both for the $(\pm)$-charge sectors and 
$(\pm)$-spin sectors. In particular, it is to be noted that
there are two possibilities
for the gap formation of the $(-)$-spin sector;
the spin gap is generated owing to locking
of the $\phi_{-\sigma}$ or $\theta_{-\sigma}$ field.
In this connection, we wish to mention that
there is a special case for which some peculiar
behavior in the mass generation is observed, i.e. the case 
of $U_1=U_2=U_3=U_4=U_5=U_6=U_7=U_8=U_9$ 
and $K_{-\sigma}=1$, where the model has 
special symmetry enhancement.\cite{fin,shu}
This situation realizes in the case with a long-range 
electron-electron interaction between parallel 
spins, $H'=V\sum_{i, s=\uparrow\downarrow}n_{i s}n_{i+1 s}$,
where the initial values of $U_8$ and $U_9$ are non-zero.
To see the special symmetry, we rewrite the interaction term $H_2$ as,
\begin{eqnarray}
H_2&=&U_1\int \frac{dx}{\alpha}[(\cos\sqrt{4\pi}\phi_{+\rho}
+\cos\sqrt{4\pi}\phi_{+\sigma} \nonumber \\
&+&\cos\sqrt{4\pi}\phi_{-\rho})
(\cos\sqrt{4\pi}\phi_{-\sigma}+\cos\sqrt{4\pi}\theta_{-\sigma}) 
\nonumber \\
&+&\cos\sqrt{4\pi}\theta_{-\rho}\cos\sqrt{4\pi}\phi_{+\sigma} 
\nonumber \\
&+&\cos\sqrt{4\pi}\theta_{-\rho}\cos\sqrt{4\pi}\phi_{+\rho} 
\nonumber \\
&+&\cos\sqrt{4\pi}\phi_{+\sigma}\cos\sqrt{4\pi}\phi_{+\rho}].
\label{int2}
\end{eqnarray}
Then the Hamiltonian is invariant under the dual transformation
$\phi_{-\sigma}\leftrightarrow\theta_{-\sigma}$.
This is nothing but the Kramers-Wannier symmetry of the Ising model.
Thus only the half part of the boson degrees of freedom 
in the $(-)$-spin sector is massive, and the other 
half-part is decoupled and forms a massless
mode of the Ising class  with the central charge
 $c=1/2$.\cite{fin,shu}
The microscopic model given by eq.(\ref{model}) does not have 
this special symmetry, so that 
we can proceed with the following analysis 
by assuming that  all the low-energy excitations are gapful.

We investigate the effects of disorder on these fixed points
in the following sections.  We note that the
effects of quenched disorder on 1D interacting electron systems
have been extensively studied before.\cite{chui,appel,suzu,gia}
Our approach is 
a generalization of these previous studies.
We introduce weak disorder potentials into the model (\ref{model}),
\begin{eqnarray}
H_{dis}&=&H_f+H_b, \label{dish} \\
H_f&=&\int dx \sum_{s=\uparrow\downarrow}
\eta_1(x)[\psi^{\dagger}_{s1 L}(x)\psi_{s1 L}(x) 
+\psi^{\dagger}_{s1 R}(x)\psi_{s1 R}(x)] \nonumber \\
&+&\int dx \sum_{s=\uparrow\downarrow}
\eta_2(x)[\psi^{\dagger}_{s2 L}(x)\psi_{s2 L}(x) 
+\psi^{\dagger}_{s2 R}(x)\psi_{s2 R}(x)] \nonumber \\
&+&\int dx \sum_{s=\uparrow\downarrow}
\eta_3(x)[e^{i(k_{\rm F1}-k_{\rm F2})x}
(\psi^{\dagger}_{s1 L}(x)\psi_{s2 L}(x) \nonumber \\
&& +\psi^{\dagger}_{s2 R}(x)\psi_{s1 R}(x))+h.c.], 
\label{for}\\
H_b&=&\int dx \sum_{s=\uparrow\downarrow}
[\xi_1(x)e^{2ik_{\rm F1}}\psi^{\dagger}_{s1 L}(x)\psi_{s1 R}(x)
+h.c.] \nonumber \\
&&+\int dx \sum_{s=\uparrow\downarrow}
[\xi_2(x)e^{2ik_{\rm F2}}\psi^{\dagger}_{s2 L}(x)\psi_{s2 R}(x)
+h.c.] \nonumber \\
&&+\int dx \sum_{s=\uparrow\downarrow}
[\xi_3(x)e^{i(k_{\rm F1}+k_{\rm F2})x}
(\psi^{\dagger}_{s1 L}(x)\psi_{s2 R}(x) \nonumber \\
&&+\psi^{\dagger}_{s2 L}(x)\psi_{s1 R}(x))+h.c.],\label{bac}
\end{eqnarray}
where $H_f$ is the forward scattering part, and
$H_b$ is the backward scattering part.
$\eta_i(x)$ and  $\xi_i(x)$ ($i=1,2$)  are the random potential 
fields within each band for forward and backward 
scatterings, respectively.  
The $\eta_3(x)$ and $\xi_3(x)$ terms represent the random hopping 
between two splitted bands.
We assume that these random fields obey 
the Gaussian distribution law,
$\langle\eta_i(x)\eta_j(x')\rangle
=D_{\eta_i}\delta_{ij}\delta(x-x')$,
$\langle\xi_i(x)\xi_j^{*}(x')\rangle
=D_{\xi_i}\delta_{ij}\delta(x-x')$,
and $D_{\eta (\xi) 1}=D_{\eta (\xi) 2}$.
We first consider the case of $D_{\eta 3}=D_{\xi 3}=0$ 
in Sec. III, IV, and V,
and take into account these random hopping terms in Sec. VI.

\section{Scaling equations for disordered systems 
in the case without random interband hopping}

In the absence of the $\eta_3$ and $\xi_3$ terms, 
we bosonize the random potential terms (\ref{for}) and (\ref{bac}),
and average over the random fields using the replica trick as in ref. 27.
Then the random potential terms in the action are given by,
\end{multicols}
\begin{eqnarray}
S_{dis}&=&-\frac{2D_{\eta +}}{\pi}\int dx\int d\tau\int d \tau'
\sum_{i,j}^n\partial_x\phi_{+\rho}^{i}(x,\tau)
\partial_x\phi_{+\rho}^{j}(x,\tau') 
-\frac{2D_{\eta -}}{\pi}\int dx\int d\tau\int d \tau'
\sum_{i,j}^n\partial_x\phi_{-\rho}^{i}(x,\tau)
\partial_x\phi_{-\rho}^{j}(x,\tau') \nonumber \\
&&-\frac{2D_{A +}}{\pi}\int dx\int d\tau\int d \tau'
\sum_{i,j}^n\frac{1}{v_{+\rho}^2}
\partial_{\tau}\phi_{+\rho}^{i}(x,\tau)
\partial_{\tau}\phi_{+\rho}^{j}(x,\tau') \nonumber \\ 
&&-\frac{2D_{A -}}{\pi}\int dx\int d\tau\int d \tau'
\sum_{i,j}^n\frac{1}{v_{-\rho}^2}
\partial_{\tau}\phi_{-\rho}^{i}(x,\tau)
\partial_{\tau}\phi_{-\rho}^{j}(x,\tau') \nonumber \\
&&-\frac{D_{\xi 1}}{\alpha^2}\int dx\int d\tau\int d \tau'
\sum_{i,j}^n\cos\sqrt{\pi}(\phi^i_{+\sigma}(x,\tau)
+\phi^i_{-\sigma}(x,\tau))
\cos\sqrt{\pi}(\phi^j_{+\sigma}(x,\tau)
+\phi^j_{-\sigma}(x,\tau)) \nonumber \\
&&\times
\cos\sqrt{\pi}(\phi^i_{+\rho}(x,\tau)+\phi^i_{-\rho}(x,\tau)
-\phi^j_{+\rho}(x,\tau')-\phi^j_{-\rho}(x,\tau')) \nonumber \\
&&-\frac{D_{\xi 2}}{\alpha^2}\int dx\int d\tau\int d \tau'
\sum_{i,j}^n\cos\sqrt{\pi}(\phi^i_{+\sigma}(x,\tau)
-\phi^i_{-\sigma}(x,\tau))
\cos\sqrt{\pi}(\phi^j_{+\sigma}(x,\tau)
-\phi^j_{-\sigma}(x,\tau)) \nonumber \\
&&\times\cos\sqrt{\pi}(\phi^i_{+\rho}(x,\tau)-\phi^i_{-\rho}(x,\tau)
-\phi^j_{+\rho}(x,\tau')+\phi^j_{-\rho}(x,\tau')),
\end{eqnarray}
\begin{multicols}{2}
\noindent
where $i,j$ are replica indices, and initially 
$D_{\eta \pm}=(D_{\eta 1}+ D_{\eta 2})/2$.
The $D_{A\pm}$ terms are generated 
in the process of renormalization,
 though initially $D_{A\pm}=0$.\cite{fuji}
The $D_{A\pm}$ term is equivalent to the interaction 
with random gauge fields,
\begin{equation}
H_{rg\pm}=-\sqrt{\frac{2}{\pi}}\int dx
A_{\pm}(x)\partial_x\theta_{\pm\rho},
\end{equation}
where the random gauge fields $A_{\pm}(x)$ obey the Gaussian 
distribution law,
$\langle A_{\alpha}(x)A_{\beta}(x')\rangle
=D_{A\alpha}\delta_{\alpha\beta}
\delta(x-x')$ ($\alpha,\beta=\pm$).
The random fields $\eta_{\pm}(x)$ and $A_{\pm}(x)$ 
can be incorporated 
into the shift of the phase fields.\cite{gia,fuji}
Defining the new fields, 
\begin{eqnarray}
\tilde{\phi}_{\pm\rho}(x)&\equiv& \phi_{\pm\rho}(x)
+\tilde{\eta}_{\pm}(x),  \\
\tilde{\eta}_{\pm}(x)&\equiv& \frac{K_{\pm\rho}}{v_{\pm\rho}}
\sqrt{\frac{2}{\pi}}
\int^x dx' \eta_{\pm}(x'), \\
\tilde{\theta}_{\pm\rho}(x)&\equiv& 
\theta_{\pm\rho}(x)+\tilde{A}_{\pm}(x),
 \\
\tilde{A}_{\pm}(x)&\equiv& 
\frac{1}{K_{\pm\rho}v_{\pm\rho}}\sqrt{\frac{2}{\pi}}
\int^x dx' A_{\pm}(x'),
\end{eqnarray} 
where $\langle\eta_{\alpha}(x)\eta_{\beta}(x')\rangle
=D_{\eta\alpha}
\delta_{\alpha\beta}\delta(x-x')$ ($\alpha,\beta=\pm$),
we can absorb the $\eta_{\pm}(x)$ and $A_{\pm}(x)$ terms into
the Gaussian part of the Hamiltonian.
Thus these terms do not affect the massless modes
apart from a minor change in correlation functions.\cite{gia,fuji}
However, as we will see momentarily, these terms are important for
suppressing the charge and spin gaps in the massive sectors. 

We now derive the scaling equations 
for the above effective Hamiltonian.
Since we have so many coupling constants, it is pretty difficult
to analyze all scaling equations even at one-loop level.
We here ignore the renormalization of velocities for simplicity.
In some previous studies\cite{kv,baf}, it was shown that
even with this approximation, the mass generation can be described
rather well qualitatively for clean systems. This is partly because
in the clean case the mass generation is mainly determined by the
scaling dimensions of the relevant operators which are not affected
by the renormalization of velocities in the lowest order. Thus this
approximation gives a qualitatively correct answer for the mass
generation in the clean case. We expect that this may be also the
case so far as we are concerned with weak disorder.
Using the operator product expansion of the U(1) Gaussian model,
and taking the replica limit $n\rightarrow 0$,
we have the scaling equations for dimensionless couplings defined by
$\tilde{U}_i=U_i/v_{+\rho}$, $\tilde{D}_{\eta,A\pm}=
D_{\eta,A\pm}/v_{\pm\rho}^2$, and 
$\tilde{D}_{\xi i}=D_{\xi i}/v_{+\rho}^2$,
\begin{eqnarray}
\frac{d \tilde{U}_1}{d l}&=&(2-\frac{1}{K_{-\rho}}
-K_{-\sigma})\tilde{U}_1 
\nonumber \\
&&-\frac{\tilde{U}_3\tilde{U}_6}{2}
-\frac{\tilde{U}_2\tilde{U}_7}{2}
-\frac{2\tilde{D}_{A-}\tilde{U}_1}{\pi^2K_{-\rho}^2}, 
\label{sc1} \\
\frac{d \tilde{U}_2}{d l}&=&(2-\frac{1}{K_{-\rho}}
-K_{+\sigma})\tilde{U}_2 
\nonumber \\
&&-\frac{\tilde{U}_3\tilde{U}_8}{2}
-\frac{\tilde{U}_1\tilde{U}_7}{2}
-\frac{2\tilde{D}_{A-}\tilde{U}_2}{\pi^2K_{-\rho}^2}, 
\label{sc2} \\
\frac{d \tilde{U}_3}{d l}&=&(2-\frac{1}{K_{-\rho}}
-K_{+\rho})\tilde{U}_3 
-\frac{\tilde{U}_1\tilde{U}_6}{2}
-\frac{\tilde{U}_5\tilde{U}_9}{2} \nonumber \\
&&-\frac{2\tilde{D}_{A-}\tilde{U}_3}{\pi^2K_{-\rho}^2}
-\frac{2\tilde{D}_{\eta +}K_{+\rho}^2\tilde{U}_3}{\pi^2}, 
\label{sc3} \\
\frac{d \tilde{U}_4}{d l}&=&(2-\frac{1}{K_{-\sigma}}
-K_{+\sigma})\tilde{U}_4
-\frac{\tilde{U}_2\tilde{U}_9}{2}
-\frac{\tilde{U}_5\tilde{U}_8}{2}, 
\label{sc4} \\
\frac{d \tilde{U}_5}{d l}&=&(2-\frac{1}{K_{-\sigma}}
-K_{+\rho})\tilde{U}_5
-\frac{\tilde{U}_4\tilde{U}_8}{2}
-\frac{\tilde{U}_3\tilde{U}_9}{2} \nonumber \\
&&-\frac{2\tilde{D}_{\eta +}K_{+\rho}^2\tilde{U}_5}{\pi^2}, 
\label{sc5} \\
\frac{d \tilde{U}_6}{d l}&=&(2-K_{-\sigma}
-K_{+\rho})\tilde{U}_6
-\frac{\tilde{U}_1\tilde{U}_3}{2}
-\frac{\tilde{U}_7\tilde{U}_8}{2} \nonumber \\
&&-\frac{2\tilde{D}_{\eta +}K_{+\rho}^2\tilde{U}_6}{\pi^2}, 
\label{sc6} \\
\frac{d \tilde{U}_7}{d l}&=&(2-K_{-\sigma}
-K_{+\sigma})\tilde{U}_7
-\frac{\tilde{U}_1\tilde{U}_2}{2}
-\frac{\tilde{U}_6\tilde{U}_8}{2} \nonumber \\
&&-\tilde{D}_{\xi 1}-\tilde{D}_{\xi 2}, \label{sc7} \\
\frac{d \tilde{U}_8}{d l}&=&(2-K_{+\sigma}
-K_{+\rho})\tilde{U}_8
-\frac{\tilde{U}_4\tilde{U}_5}{2}
-\frac{\tilde{U}_6\tilde{U}_7}{2} \nonumber \\
&&-\frac{2\tilde{D}_{\eta +}K_{+\rho}^2\tilde{U}_8}{\pi^2}, 
\label{sc8} \\
\frac{d \tilde{U}_9}{d l}&=&(2-\frac{1}{K_{-\rho}}
-\frac{1}{K_{-\sigma}})
\tilde{U}_9
-\frac{\tilde{U}_2\tilde{U}_4}{2}
-\frac{\tilde{U}_3\tilde{U}_5}{2} \nonumber \\
&&-\frac{2\tilde{D}_{A-}\tilde{U}_9}{\pi^2K_{-\rho}^2}, 
\label{sc9} \\
\frac{d K_{+\rho}}{d l}&=&-\frac{\pi K_{+\rho}^2}{2}
[\tilde{U}_3^2+\tilde{U}_5^2+\tilde{U}_6^2
+\tilde{U}_8^2] \nonumber \\
&&-\frac{\pi K_{+\rho}^2}{16}(\tilde{D}_{\xi 1}+\tilde{D}_{\xi 2}), 
\label{sc10} \\
\frac{d K_{-\rho}}{d l}&=&\frac{\pi}{2}
[\tilde{U}_1^2+\tilde{U}_2^2+\tilde{U}_3^2+\tilde{U}_9^2] 
\nonumber \\
&&-\frac{\pi K_{-\rho}^2}{16}(\tilde{D}_{\xi 1}+\tilde{D}_{\xi 2}), 
\label{sc11} \\
\frac{d K_{+\sigma}}{d l}&=&-\frac{\pi K_{+\sigma}^2}{2}
[\tilde{U}_2^2+\tilde{U}_4^2+\tilde{U}_7^2
+U_8^2] \nonumber \\
&&-\frac{\pi K_{+\sigma}^2}{16}(\tilde{D}_{\xi 1}+\tilde{D}_{\xi 2}), 
\label{sc12} \\
\frac{d K_{-\sigma}}{d l}&=&\frac{\pi}{2}
[\tilde{U}_4^2+\tilde{U}_5^2+\tilde{U}_9^2] 
-\frac{\pi K_{-\sigma}^2}{2}
[\tilde{U}_1^2+\tilde{U}_6^2+\tilde{U}_7^2] \nonumber \\
&-&\frac{\pi K_{-\sigma}^2}{16}(\tilde{D}_{\xi 1}+\tilde{D}_{\xi 2}), 
\label{sc13} \\
\frac{d \tilde{D}_{\eta\pm}}{d l}&=&\tilde{D}_{\eta \pm}
+\pi^2(\tilde{D}_{\xi 1}^2+\tilde{D}_{\xi 2}^2), 
\label{sc16} \\
\frac{d \tilde{D}_{A\pm}}{d l}&=&\tilde{D}_{A \pm}
+\pi^2(\tilde{D}_{\xi 1}^2+\tilde{D}_{\xi 2}^2), 
\label{sc17} \\
\frac{d \tilde{D}_{\xi i}}{d l}&=&
(3-\frac{K_{+\sigma}}{2}-\frac{K_{-\sigma}}{2}
-\frac{K_{+\rho}}{2}-\frac{K_{-\rho}}{2})\tilde{D}_{\xi i} 
\nonumber \\
&&-\tilde{U}_7\tilde{D}_{\xi i}, \qquad i=1,2. \label{sc18} 
\end{eqnarray}
In the absence of disorder, $D_{\eta\pm}=D_{A\pm}=D_{\xi i}=0$,
the excitation gaps open 
above the ground state energy in the $(\pm)$-charge and 
($\pm$)-spin sectors.
We have two possibilities in the massive 
phase of the $(-)$-spin sector
as mentioned in Sec. II.
If $\tilde{U}_1$, $\tilde{U}_6$, 
and $\tilde{U}_7$ scale to strong-coupling 
regime, and the $\tilde{U}_4$, $\tilde{U}_5$, and $\tilde{U}_9$ terms
are irrelevant, the phase field $\phi_{-\sigma}$ is locked, 
and the correlations 
for the operator $\exp(ia\theta_{-\sigma})$ shows exponential decay. 
We refer to this case as {\it case A}.
To the contrary, if the $\tilde{U}_4$, $\tilde{U}_5$, 
and $\tilde{U}_9$ terms are relevant, 
and the $\tilde{U}_1$, $\tilde{U}_6$, and $\tilde{U}_7$ terms are
irrelevant, $\theta_{-\sigma}$ is locked.
We refer to this case as {\it case B}. 
In the subsequent sections, we discuss the effects of disorder
on these two kinds of the fixed points.

\section{Disordered fixed point:  {\it case A}}

Here we consider the effects of quenched disorder 
on the fixed point of {\it case A},  where
the phase field $\phi_{-\sigma}$ is locked, 
and  $\exp(ia\theta_{-\sigma})$ is the disorder
parameter. 
In this case, the $\tilde{U}_1$, $\tilde{U}_2$, $\tilde{U}_3$, 
$\tilde{U}_6$, $\tilde{U}_7$, and $\tilde{U}_8$ 
terms are relevant, and the $\tilde{U}_4$, $\tilde{U}_5$, 
$\tilde{U}_9$ terms are irrelevant in the clean system. 
The sufficiently strong forward scattering due to impurities
$\tilde{D}_{\eta +}$
may suppress the mass-generating interactions, 
$\tilde{U}_3$, $\tilde{U}_6$, 
and $\tilde{U}_8$, as seen from 
Eqs.(\ref{sc3}), (\ref{sc6}), and (\ref{sc8}). 
It turns out from Eqs.(\ref{sc1}) and (\ref{sc2}) 
that this is not the case for 
the $\tilde{U}_1$ and $\tilde{U}_2$ terms which
bear the charge gap in the $(-)$-sector, because they
are not coupled with the random forward scattering.
We note, however, that these terms should be
 suppressed in the presence of  sufficiently strong random 
gauge fields $A_{-}$ generated 
by random backward scattering due to impurities.
Therefore in order to close the charge gap in the $(-)$-sector, 
one needs the random backward scattering due to impurities.
This point makes a clear contrast to the case of a single chain model
where the Mott-Hubbard gap is collapsed only by the random forward
scattering due to impurities.\cite{fuji} 

Note that the $\tilde{U}_7$ term which generates the spin gap 
in $\phi_{\pm\sigma}$ fields
is not affected by the random forward scattering, because
the random forward scattering couples only with
 the charge degrees of freedom. 
On the other hand,
it is seen from Eq.(\ref{sc7}) that 
the backward scattering due to impurities $\tilde{D}_{\xi i}$ 
may drive $\tilde{U}_7$ to
a large negative value.
However, if the initial values of $\tilde{D}_{\xi i}$ 
are small enough
compared to that of $\tilde{U}_7$, the spin 
gaps of $\phi_{\pm\sigma}$ fields 
generated by the $\tilde{U}_7$ term may survive.
Based on this observation, we start with the limiting case 
$\tilde{U}_7\rightarrow +\infty$ for simplicity.
Then after the spin gap formation, we can eliminate
the spin degrees of freedom from the scaling equations.
Suppressing irrelevant terms, 
we have the reduced scaling equations,
\begin{eqnarray}
\frac{d \tilde{U}_1}{d l}&=&(2-\frac{1}{K_{-\rho}})\tilde{U}_1
-\frac{\tilde{U}_3\tilde{U}_6}{2}
-\frac{2\tilde{D}_{A-}\tilde{U}_1}{\pi^2K_{-\rho}^2}, 
\label{sca1} \\
\frac{d \tilde{U}_2}{d l}&=&(2-\frac{1}{K_{-\rho}})\tilde{U}_2
-\frac{\tilde{U}_3\tilde{U}_8}{2}
-\frac{2\tilde{D}_{A-}\tilde{U}_2}{\pi^2K_{-\rho}^2},             
\label{sca2} \\
\frac{d \tilde{U}_3}{d l}&=&
(2-\frac{1}{K_{-\rho}}-K_{+\rho})\tilde{U}_3
-\frac{\tilde{U}_1\tilde{U}_6}{2}-\frac{\tilde{U}_5\tilde{U}_9}{2} 
\nonumber \\
&&-\frac{2\tilde{D}_{A-}\tilde{U}_3}{\pi^2K_{-\rho}^2}
-\frac{2\tilde{D}_{\eta +}K_{+\rho}^2\tilde{U}_3}{\pi^2}, 
\label{sca3} \\
\frac{d \tilde{U}_6}{d l}&=&(2-K_{+\rho})\tilde{U}_6
-\frac{\tilde{U}_1\tilde{U}_3}{2}
-\frac{2\tilde{D}_{\eta +}K_{+\rho}^2\tilde{U}_6}{\pi^2}, 
\label{sca6} \\
\frac{d \tilde{U}_8}{d l}&=&(2-K_{+\rho})\tilde{U}_8
-\frac{2\tilde{D}_{\eta +}K_{+\rho}^2\tilde{U}_8}{\pi^2}, 
\label{sca8} \\
\frac{d K_{+\rho}}{d l}&=&-\frac{\pi K_{+\rho}^2}{2}
[\tilde{U}_3^2+\tilde{U}_6^2
+\tilde{U}_8^2] \nonumber \\
&&-\frac{\pi K_{+\rho}^2}{16}(\tilde{D}_{\xi 1}+\tilde{D}_{\xi 2}),
\label{sca10} \\
\frac{d K_{-\rho}}{d l}&=&\frac{\pi}{2}
[\tilde{U}_1^2+\tilde{U}_2^2+\tilde{U}_3^2]
-\frac{\pi K_{-\rho}^2}{16}(\tilde{D}_{\xi 1}+\tilde{D}_{\xi 2}),
\label{sca11} \\
\frac{d \tilde{D}_{\xi i}}{d l}&=&
(3-\frac{K_{+\rho}}{2}-\frac{K_{-\rho}}{2})\tilde{D}_{\xi i}, 
\qquad i=1,2. \label{sca18}
\end{eqnarray}
Since initially $\tilde{D}_{\eta\pm}>\tilde{D}_{A\pm}=0$,  
and they
obey the same scaling equations (\ref{sc16}) and (\ref{sc17}),
$\tilde{D}_{\eta\pm}$ flows into strong-coupling regime 
faster than $\tilde{D}_{A\pm}$.
Thus there exists a parameter region where 
the $\tilde{U}_3$, $\tilde{U}_6$, and $\tilde{U}_8$ terms 
are suppressed for sufficiently 
large $\tilde{D}_{\eta +}$, though the $\tilde{U}_1$ 
and $\tilde{U}_2$ terms are still
relevant.
Then the charge gap in $(+)$-charge mode is collapsed.
Since the $\theta_{-\rho}$ field is locked due to the $\tilde{U}_1$ 
and $\tilde{U}_2$
terms, $K_{-\rho}$ scales to a large value, and 
the $\tilde{D}_{\xi i}$ terms are suppressed, 
as seen from Eq.(\ref{sca18}).
However, these terms are proved to be relevant, 
after one integrates out the $\phi_{-\rho}$ field, 
following the argument by Orignac and Giamarchi\cite{og}.
as in the case away from half-flling.\cite{og}
Then the $(+)$-charge mode is in the Anderson localized state,
while the $(-)$-charge mode is still in the Mott insulating state.

As the value of $\tilde{D}_{A-}$ becomes larger, 
the $\tilde{U}_1$, $\tilde{U}_2$
terms are suppressed, and all excitation gaps are collapsed.
Then, the backward scattering due to impurities, $\tilde{D}_{\xi i}$
is relevant, and consequently the Anderson localization occurs.
$\tilde{U}_7$ flows to a large negative value 
because of the $\tilde{D}_{\xi i}$ term,
as seen from Eq.(\ref{sc7}).
Thus the spin degrees of freedom are frozen.
The fixed point is identified with the pinned CDW state
which is similar to that found 
in the disordered single chain model.\cite{gia}

In order to confirm the above arguments, 
we solve the scaling equations
(\ref{sc1})-(\ref{sc18}) numerically for several initial values
of coupling constants.
The numerical results for the flow of 
coupling constants are shown in FIG.1. 
In order to realize the massive spin state due to the locking of 
$\phi_{-\sigma}$ in the clean system, 
we have chosen the initial value of the TL liquid
parameter in the $(-)$-spin sector as $K_{-\sigma}(0)=0.86$.
The other TL parameters are set to be
$K_{\pm\rho}(0)=0.98$ and $K_{+\sigma}(0)=1.0$.
For these initial values, the system definitely flows toward  
the low-energy fixed point of {\it case A}
in the absence of disorder.
In FIG.1(a) the results in the case 
without quenched disorder are shown.
$\tilde{U}_1$, $\tilde{U}_2$, $\tilde{U}_3$, $\tilde{U}_6$, 
$\tilde{U}_7$, and $\tilde{U}_8$ flow into a 
strong-coupling regime, while 
$\tilde{U}_4$, $\tilde{U}_5$, and $\tilde{U}_9$ scale to $0$.
As a consequence, the phase fields
$\phi_{+\rho}$, $\theta_{-\rho}$. $\phi_{\pm\sigma}$ 
are locked. This fixed point 
is stable against the disorder 
weaker than mass-generating interactions, because of
the presence of excitation gaps in all modes.
As the quenched disorder becomes stronger, 
some different fixed points
manifest themselves. 
In FIG.1(b), we show the results
for the initial values 
$\tilde{D}_{\eta\pm}(0)=0.9$ and $\tilde{D}_{\xi i}(0)=0.005$.   
Although  $\tilde{U}_1$, $\tilde{U}_2$ and $\tilde{U}_7$
still scale to a strong-coupling regime,
$\tilde{U}_3$, $\tilde{U}_6$, and $\tilde{U}_8$ scale to $0$.
As a result, the charge gap
in the $(+)$-mode is collapsed.
Since $\tilde{D}_{\xi i}$ is relevant,
the charge mode in the $(+)$-sector is localized, though
the charge mode in the $(-)$-sector is still massive.
Thus the Anderson localized state and the Mott 
insulating state coexist in this phase.
Since $\tilde{U}_7\rightarrow -\infty$,
the spin degrees of freedom are frozen, and may be in
the random spin singlet state. 
In FIG.1(c),
the results for $D_{\eta\pm}(0)=0.9$ 
and $D_{\xi i}(0)=0.01$ are shown. 
In this case, all mass-generating interactions except 
$\tilde{U}_7$ in Eq.(\ref{int1}) 
flow to $0$. The charge gaps are thus collapsed in this phase.
Since $\tilde{D}_{\xi i}$ scales to a strong-coupling regime,
the Anderson localized state realizes.
The relevance of the $\tilde{U}_7$ term implies 
freezing of spin degrees of freedom
in the $(\pm)$-sectors. 

In summary, we have found three kinds of fixed points
in the presence of quenched disorder:
(i) the clean fixed point where the disorder effect 
is entirely irrelevant, 
(ii) Anderson-insulator (I) where
the $(+)$-charge mode is in the Anderson localized state, 
the $(-)$-charge mode is in the Mott insulating state, 
and the spin degrees of freedom are massive or frozen,
(iii) Anderson-insulator (II) where
all charge modes are in the Anderson localized state, and
spin degrees of freedom are frozen, {\it i.e.} pinned CDW.
We thus end up with the schematic phase 
diagram for charge degrees 
of freedom as shown in FIG.2.

\section{Disordered fixed point: {\it case B}}

Here we consider the effects of disorder on 
the fixed point of {\it case B}:
$\theta_{-\sigma}$ is locked, and
$\exp(ia\phi_{-\sigma})$ is the disorder parameter field 
in the absence of quenched disorder.
In this case, the $\tilde{U}_2$, $\tilde{U}_3$, $\tilde{U}_4$, 
$\tilde{U}_5$, $\tilde{U}_8$, and $\tilde{U}_9$ terms
are relevant in the clean system.
We see from Eq.(\ref{sc13}) that $K_{-\sigma}$ is decreased by 
the backward scattering due to impurities $\tilde{D}_{\xi i}$.
Thus it suppresses the growth of $\tilde{U}_4$, $\tilde{U}_5$, 
and $\tilde{U}_9$
and prevents the spin gap formation due to locking of
the $\theta_{-\sigma}$ field. 
Moreover it follows from Eq.(\ref{sc7}) that 
if the $\tilde{D}_{\xi i}$ term is
relevant, $\tilde{U}_7$ flows to a large negative  value,
resulting in freezing of the spin degrees of freedom,
and the suppression of the $\tilde{U}_4$ term as in Sec. IV.
However for sufficiently small $\tilde{D}_{\xi i}$, 
the $\tilde{U}_4$ term
is relevant and the spin gap opens 
owing to locking of $\theta_{-\sigma}$.
After the generation of the spin gap due to locking of
the $\theta_{-\sigma}$ and
$\phi_{+\sigma}$ fields, we can eliminate 
the spin degrees of freedom,
and obtain the scaling equations for 
the charge degrees of freedom,
\begin{eqnarray}
\frac{d \tilde{U}_2}{d l}&=&(2-\frac{1}{K_{-\rho}})\tilde{U}_2
-\frac{\tilde{U}_3\tilde{U}_8}{2}
-\frac{2\tilde{D}_{A-}\tilde{U}_2}{\pi^2K_{-\rho}^2}, 
\label{scb2} \\
\frac{d \tilde{U}_3}{d l}&=&(2-\frac{1}{K_{-\rho}}
-K_{+\rho})\tilde{U}_3
-\frac{\tilde{U}_5\tilde{U}_9}{2} \nonumber \\
&&-\frac{2\tilde{D}_{A-}\tilde{U}_3}{\pi^2K_{-\rho}^2}
-\frac{2\tilde{D}_{\eta +}K_{+\rho}^2\tilde{U}_3}{\pi^2}, 
\label{scb3} \\
\frac{d \tilde{U}_5}{d l}&=&(2-K_{+\rho})\tilde{U}_5
-\frac{\tilde{U}_3\tilde{U}_9}{2} \nonumber \\
&&-\frac{2\tilde{D}_{\eta +}K_{+\rho}^2\tilde{U}_5}{\pi^2}, 
\label{scb5} \\
\frac{d \tilde{U}_8}{d l}&=&(2-K_{+\rho})\tilde{U}_8
-\frac{2\tilde{D}_{\eta +}K_{+\rho}^2\tilde{U}_8}{\pi^2}, 
\label{scb8} \\
\frac{d \tilde{U}_9}{d l}&=&(2-\frac{1}{K_{-\rho}})\tilde{U}_9
-\frac{\tilde{U}_3\tilde{U}_5}{2} 
-\frac{2\tilde{D}_{A-}\tilde{U}_9}{\pi^2K_{-\rho}^2}, 
\label{scb9} \\
\frac{d K_{+\rho}}{d l}&=&-\frac{\pi K_{+\rho}^2}{2}
[\tilde{U}_3^2+\tilde{U}_5^2
+\tilde{U}_8^2], 
\nonumber \\
&&-\frac{\pi K_{+\rho}^2}{16}(\tilde{D}_{\xi 1}
+\tilde{D}_{\xi 2}), \label{scb10} \\
\frac{d K_{-\rho}}{d l}&=&\frac{\pi}{2}
[\tilde{U}_2^2+\tilde{U}_3^2+\tilde{U}_9^2]. 
\nonumber \\
&&-\frac{\pi K_{-\rho}^2}{16}(\tilde{D}_{\xi 1}+\tilde{D}_{\xi 2}), 
\label{scb11} \\
\frac{d \tilde{D}_{\xi 1,2}}{d l}&=&(3-2K_{+\rho}
-2K_{-\rho})\tilde{D}_{\xi i}. \label{scb12} 
\end{eqnarray}
As pointed out in Sec. IV, $\tilde{D}_{\eta\pm}$ 
grows faster than $\tilde{D}_{A\pm}$.
Thus there may exist a parameter region where
$\tilde{D}_{\eta +}$ suppresses $\tilde{U}_3$, $\tilde{U}_5$, 
and $\tilde{U}_8$,
but the $\tilde{U}_2$ and $\tilde{U}_9$ terms are still relevant,
bearing the mass gap in the $(-)$-charge mode.
In this parameter region, only the charge gap in the 
$(+)$-charge sector is collapsed.
The low-energy properties of this sector is determined by
the $\tilde{D}_{\xi i}$-terms.
After integrating out the massive $(-)$-charge mode\cite{og},
we find that the scaling equation for 
the $\tilde{D}_{\xi 1,2}$-terms are changed to,
\begin{equation}
\frac{d \tilde{D}_{\xi 1,2}}{d l}=(3-8K_{+\rho})\tilde{D}_{\xi i}.
\end{equation}
Thus the impurity backward scattering is suppressed 
in comparison with
single chain systems.
For $K_{+\rho}>3/8$, the $D_{\xi 1,2}$-terms are irrelevant.
However it does not directly mean the occurrence of delocalization,
since we should actually include the random hopping between two bands, 
$D_{\eta 3}$, $D_{\xi 3}$.
In the case of $D_{\eta 3}\neq 0$ and $D_{\xi 3}\neq 0$,
the CDW order couples directly to the $D_{\xi 3}$ term,
and then the pinning of the CDW occurs for infinitesimally small
$D_{\xi 3}$ as will be discussed in Sec. VI.
Therefore the $(+)$-charge mode is in the Anderson localized state
with the vanishing Drude weight and the non-zero charge susceptibility.
For a sufficiently large value of $\tilde{D}_{A-}$, the charge gap
in the $(-)$-mode may be also collapsed.
All the charge modes are in the Anderson localized state.

In order to specify the low-energy fixed points correctly,
we have solved the full scaling equations 
(\ref{sc1})-(\ref{sc18}) 
numerically for some initial values of couplings.
The numerical results are shown in FIG.3.
In FIG.3(a), we display the results for clean systems.
In order to realize the spin gapped state due to locking of 
$\theta_{-\sigma}$, 
we choose the initial values of the parameters 
$K_{-\sigma}(0)=1.1$, $K_{+\sigma}(0)=1.0$, and 
$K_{\pm\rho}(0)=0.98$.
For these parameters,
$\tilde{U}_2$, $\tilde{U}_3$, $\tilde{U}_4$, 
$\tilde{U}_5$, $\tilde{U}_8$, 
and $\tilde{U}_9$ scale to 
a strong-coupling regime, and $\tilde{U}_1$, $\tilde{U}_6$, and 
$\tilde{U}_7$ scale to $0$.
Then the phase fields $\phi_{+\rho}$, 
$\theta_{-\rho}$, $\phi_{+\sigma}$, 
and $\theta_{-\sigma}$ are locked.
This fixed point with the spin and charge gaps 
is stable as far as
disorder is much weaker than mass-generating interactions.
In FIG.3(b), the results for $\tilde{D}_{\eta\pm}(0)=0.9$,
and $\tilde{D}_{\xi i}(0)=0.007$ are shown.
This flow ends up at the same fixed point as that shown in FIG.1(b),
where the backward scattering due to impurities is relevant, 
and thus
the Anderson localization occurs in the $(+)$-charge mode, 
though the $(-)$-charge mode is 
in the Mott insulating state.
The spin degrees of freedom are also massive or frozen.
Note that the freezing of the spin degrees of freedom 
in the $(-)$-sector
is not due to locking of $\theta_{-\sigma}$ 
but due to $\phi_{-\sigma}$.
Thus the fixed point properties of the $(-)$-spin sector are 
different from those for the clean case where locking of
$\theta_{-\sigma}$ occurs. 
In FIG.3(c), we show the results for $\tilde{D}_{\eta\pm}(0)=0.9$ 
and $\tilde{D}_{\xi i}(0)=0.01$.
In this case, the flows of coupling 
constants are qualitatively the same
as those shown in FIG.1(c); {\it i.e.}
all excitation gaps are collapsed, and the Anderson localized state
with frozen spin degrees of freedom is realized. 
Thus for sufficiently strong impurity scattering, 
the disordered fixed point
of {\it case B} flows into the same class as that of {\it case A}.
In summary, we have found three kinds of fixed points as in 
{\it case A}. 
The schematic phase diagram for the charge degrees of freedom 
is also given by FIG.2.

\section{Effects of random hopping between two bands}

We now take into account the effects of 
random hopping between two bands,
$\eta_3$ and $\xi_3$.
First we consider only random forward scattering, and omit
random backward scattering for a while.
Going back to the fermion representation,
we write down the kinetic energy part and 
the impurity forward scattering part of the Hamiltonian,
\begin{eqnarray}
H_0'&=&\int dx \sum_{s=\uparrow\downarrow}
[v_1(\psi^{\dagger}_{s1 L}\partial_x\psi_{s1 L}
-\psi^{\dagger}_{s1 R}\partial_x\psi_{s1 R}) \nonumber \\
&+&v_2(\psi^{\dagger}_{s2 L}\partial_x\psi_{s2 L}
-\psi^{\dagger}_{s2 R}\partial_x\psi_{s2 R})] \nonumber \\
&+&\int dx \sum_{s=\uparrow\downarrow}
\eta_1(x)[\psi^{\dagger}_{s1 L}(x)\psi_{s1 L}(x) 
+\psi^{\dagger}_{s1 R}(x)\psi_{s1 R}(x)]
\nonumber \\
&+&\int dx \sum_{s=\uparrow\downarrow}
\eta_2(x)[\psi^{\dagger}_{s2 L}(x)\psi_{s2 L}(x) 
+\psi^{\dagger}_{s2 R}(x)\psi_{s2 R}(x)]
\nonumber \\
&&+\int dx \sum_{s=\uparrow\downarrow}
\eta_3(x)[e^{i(k_{\rm F1}-k_{\rm F2})x}
(\psi^{\dagger}_{s1 L}(x)\psi_{s2 L}(x) \nonumber \\
&& +\psi^{\dagger}_{s2 R}(x)\psi_{s1 R}(x))+h.c.].
\label{h06}
\end{eqnarray}
In the case of half-filling, $v_1=v_2\equiv v$ holds.
Defining the spinor fields,
\begin{equation}
\psi_{s L(R)}=\left(
\begin{array}{c}
\psi_{s1 L(R)} \\ \psi_{s2 L(R)}
\end{array}\right),
\end{equation}
we cast the Hamiltonian Eq.(\ref{h06}) into 
the following form,
\begin{eqnarray}
H_0'&=&\int dx\sum_{s=\uparrow\downarrow}
v(\psi^{\dagger}_{s L}\partial_x\psi_{s L}
-\psi^{\dagger}_{s R}\partial_x\psi_{s R}) \nonumber \\
&+&\int  dx\sum_{s=\uparrow\downarrow}
\psi^{\dagger}_{s L}
\Bigl(\frac{\eta_1+\eta_2}{2}{\bf 1}+\frac{\eta_1-\eta_2}{2}\tau^z
\nonumber \\
&+&\eta_3 e^{i(k_{\rm F1}-k_{\rm F2})x}\frac{\tau^{+}}{2}
+\eta_3^{*} e^{-i(k_{\rm F1}-k_{\rm F2})x}\frac{\tau^{-}}{2}\Bigr)
\psi_{s L} \nonumber \\
&+&\int  dx\sum_{s=\uparrow\downarrow}
\psi^{\dagger}_{s R}
\Bigl(\frac{\eta_1+\eta_2}{2}{\bf 1}+\frac{\eta_1-\eta_2}{2}\tau^z
\nonumber \\
&+&\eta_3^{*} e^{-i(k_{\rm F1}-k_{\rm F2})x}\frac{\tau^{+}}{2}
+\eta_3 e^{i(k_{\rm F1}-k_{\rm F2})x}\frac{\tau^{-}}{2}\Bigr)
\psi_{s R},
\end{eqnarray}
where {\bf 1} is a $2\times2$ unit matrix, 
and $\tau^z$, $\tau^{\pm}$
are the Pauli matrices. 
This Hamiltonian can be diagonalized by using the unitary 
transformations defined by,
\begin{eqnarray}
\tilde{\psi}_{s L}&=&U_L\psi_{s L}, \\
U_L&\equiv&\exp\Bigl(-\frac{i}{v}\int^x_{-\infty}dx'
\Bigl[\frac{\eta_1(x')+\eta_2(x')}{2}
{\bf 1} \nonumber \\
&&+\frac{\eta_1(x')-\eta_2(x')}{2}\tau^z \nonumber \\ 
&&+\eta_3(x')e^{i(k_{\rm F1}-k_{\rm F2})x'}\frac{\tau^{+}}{2} 
\nonumber \\
&&+\eta_3^{*}(x')e^{-i(k_{\rm F1}-k_{\rm F2})x'}
\frac{\tau^{-}}{2}\Bigr]
\Bigr), \label{uni}
\end{eqnarray}
and $U_R=U_L^{\dagger}$.
We thus end up with the diagonalized form of $H_0'$,
\begin{equation}
H_0'=\int dx \sum_sv(\tilde{\psi}^{\dagger}_{s L}
\partial_x\tilde{\psi}_{s L}
-\tilde{\psi}^{\dagger}_{s R}\partial_x\tilde{\psi}_{s R}).
\end{equation}
Therefore the forward scattering part of the random hopping
can be incorporated into the free part of the Hamiltonian.
Under the unitary transformation (\ref{uni}),
the forward scattering terms of electron-electron interaction are
invariant.
The backward and umklapp scattering terms of electron-electron
interaction (\ref{int1}) which may generate the mass gaps are 
multiplied with
exponentially decaying factors by this transformation. 
Thus for sufficiently strong impurity forward 
scattering both of the charge and spin gaps
collapse as in the previous sections.

Let us now take into account random backward scattering.
The backward scattering term of random hopping is expressed as
\begin{equation}
H_{\xi3}=\int dx [\xi_3(x)e^{i(k_{\rm F1}+k_{\rm F2})x}
O_{CDW}(x)+h.c.].
\end{equation}
The scaling equation for $D_{\xi 3}$ is given by
\begin{eqnarray}
\frac{d \tilde{D}_{\xi 3}}{d l}&=&\Bigl(3-\frac{K_{+\rho}}{2}
-\frac{1}{2K_{-\rho}}
-\frac{K_{+\sigma}}{2}-\frac{1}{2K_{-\sigma}} \nonumber \\
&&-\tilde{U}_4\Bigr)
\tilde{D}_{\xi 3},\label{s3}
\end{eqnarray}
where $\tilde{D}_{\xi 3}=D_{\xi 3}/v_{+\rho}^2$.
It is seen that the $D_{\xi 3}$ term couples with the $U_4$ term.
Then the scaling equation (\ref{sc4}) is altered to
\begin{eqnarray}
\frac{d \tilde{U}_4}{d l}&=&
\Bigl(2-K_{+\sigma}-\frac{1}{K_{-\sigma}}\Bigr)
\tilde{U}_4-\tilde{U}_2\tilde{U}_9-\tilde{U}_5\tilde{U}_8 
\nonumber \\
&&-\tilde{D}_{\xi 3}.
\end{eqnarray} 
Therefore in the case that the mass gaps open due to locking of
the $\phi_{+\rho}$,
$\theta_{-\rho}$, $\phi_{+\sigma}$, and $\theta_{-\sigma}$ fields
in the clean system ({\it case B}),
$3-K_{+\rho}/2-1/2K_{-\rho}
-K_{+\sigma}/2-1/2K_{-\sigma}-\tilde{U}_4>0$
holds, and the $D_{\xi 3}$ term is relevant. Then,
the Anderson localization inevitably occurs in the $(+)$-charge 
sector provided that the impurity forward scattering 
suppresses the charge gap as mentioned in Sec. V. 
This fixed point is the pinned CDW.
If the random backward scattering $\tilde{D}_{\xi 1,2}$ 
drives $\tilde{U}_7$ to $-\infty$,
the $\phi_{-\sigma}$ field is locked.
Then in the presence of the gap in the $(-)$-charge 
sector, integrating out the $\theta_{-\rho}$ field 
as in ref. 16, we find that the scaling equation (\ref{s3}) 
is changed to,
\begin{equation}
\frac{d \tilde{D}_{\xi 3}}{d l}=(3-2K_{+\rho})\tilde{D}_{\xi 3}.
\label{s32}
\end{equation} 
Thus although the $\tilde{D}_{\xi 3}$-term is suppressed,
it is still relevant for the repulsive interaction $K_{+\rho}<1$,
and the Anderson localization occurs in the $(+)$-charge sector.

If the spin gap opens owing to locking of 
the $\phi_{-\sigma}$ field in the clean system
({\it case A}), the scaling equation for $\tilde{D}_{\xi 3}$
is also given by Eq.(\ref{s32}).
Then the random hopping between two bands enhances
the Anderson localization obtained in Sec. IV.

\section{Conclusion}

We have  systematically studied the effects of quenched disorder
on the spin-gapped state of two chain coupled Hubbard ladder
model at half-filling.
We have shown that 
a sufficiently strong forward scattering
due to impurities and the random gauge fields generated by 
impurity backward scattering collapse
all the charge gaps as well as the spin gaps,
and consequently the Anderson localization 
takes place in all the charge sectors 
owing to the impurity backward scattering.
We have also found the phase that  one charge mode is 
in the Anderson localized state, while
the other one is in the massive state.

This work was partly supported by a Grant-in-Aid from the Ministry
of Education, Science and Culture.

\begin{figure}
\centerline{\epsfxsize=6.5cm \epsfbox{flowa1.eps}}
{FIG 1. Plots of $\tilde{U}_i$ and $\tilde{D}_{\xi}$ vs 
the scaling parameter $l$ in the case A
for different initial values of $\tilde{D}_{\eta\pm}$ 
and $\tilde{D}_{\xi i}$:
(a) $\tilde{D}_{\eta\pm}(0)=\tilde{D}_{\xi i}(0)=0$.
(b) $\tilde{D}_{\eta\pm}(0)=0.9$,
$\tilde{D}_{\xi i}(0)=0.005$. (c) $\tilde{D}_{\eta\pm}(0)=0.9$,
$\tilde{D}_{\xi i}(0)=0.01$.
The initial values of the TL liquid parameters
and $\tilde{U}_i$ are set as
$K_{\pm\rho}(0)=0.98$, $K_{+\sigma}(0)=1.0$, 
$K_{-\sigma}(0)=0.86$,
and $\tilde{U}_i(0)=0.1$ for $i=1\sim 7$, 
$\tilde{U}_8(0)=\tilde{U}_9(0)=0$. 
As the initial values of $\tilde{D}_{\eta}$ 
and $\tilde{D}_{\xi i}$ 
increases, the mass generating terms scale to zero,
and the Anderson localized state realizes.}
\end{figure}
\begin{figure}
\centerline{\epsfxsize=6.8cm \epsfbox{lasphase.eps}}
{FIG 2. Phase diagram in the $D_{\eta\pm}$-$D_{\xi i}$ plane.
(i) Mott insulator (MI) in both the $(+)$ and $(-)$-charge sectors.
(ii) Anderson localization (AL)
in the $(+)$-charge sector, and  MI in the $(-)$-charge sector.
(iii) AL in both the $(+)$ and $(-)$-charge sectors.}
\end{figure}
\begin{figure}
\centerline{\epsfxsize=6.5cm \epsfbox{flowb.eps}}
{FIG 3. Plots
of $\tilde{U}_i$ and $\tilde{D}_{\xi}$ vs $l$ in the case B.
for the different initial values of $\tilde{D}_{\eta\pm}$
and $\tilde{D}_{\xi i}$:
(a) $\tilde{D}_{\eta\pm}(0)=\tilde{D}_{\xi i}(0)=0$.
(b) $\tilde{D}_{\eta\pm}(0)=0.9$,
$\tilde{D}_{\xi i}(0)=0.007$. (c) $\tilde{D}_{\eta\pm}(0)=0.9$,
$\tilde{D}_{\xi i}(0)=0.01$.
The initial values of the TL parameters and $\tilde{U}_i$
are set as $K_{\pm\rho}(0)=0.98$, $K_{+\sigma}(0)=1.0$, 
$K_{-\sigma}(0)=1.1$,
$\tilde{U}_i(0)=0.1$ for $i=1\sim 7$, 
and $\tilde{U}_8(0)=\tilde{U}_9(0)=0$. 
}
\end{figure}

\end{multicols}

\end{document}